\documentclass[pra,twocolumn,showpacs,floatfix]{revtex4}

\usepackage{natbib}
\usepackage{graphicx}
\usepackage{dcolumn}
\usepackage{amsmath}
\usepackage{amssymb}

\def\beq{\begin{equation}}
\def\eeq{\end{equation}}

\def\om{\omega}

\def\eps{\epsilon}
\def\veps{\varepsilon}

\def\s{\sigma}
\def\D{\Delta}

\def\ad{a^\dagger}

\def\rd{{\rm{d}}}

\def\p{\phi}
\def\bp{\bar{\psi}}

\def\ra{\rightarrow}

\def\Cc{\mathbb{C}}
\def\Zz{\mathbb{Z}}
\def\R{\mathbb{R}}

\def\dd{\frac{\rd}{\rd z}}

\def\Hp{{\cal H}_+}
\def\Hm{{\cal H}_-}
\def\ua{\uparrow}
\def\da{\downarrow}
\def\dd{\frac{\rd}{\rd z}}

\begin{document}
\title{On the Integrability of the Rabi Model}

\author{ D.~Braak\footnote{email: Daniel.Braak@physik.uni-augsburg.de}
}
\affiliation{EP VI and Center for Electronic Correlations and Magnetism, 
University of Augsburg, 86135 Augsburg, Germany}

\date{\today}

\begin{abstract}
The exact spectrum  of the Rabi hamiltonian
is analytically found for arbitrary coupling strength and detuning.
I present a criterion for integrability of 
quantum systems containing
discrete degrees of freedom which shows that in this case 
a finite symmetry group may entail
integrability, even without the presence of conserved charges
beyond the hamiltonian itself.
Moreover, I introduce and solve a natural generalization of the Rabi
model
which has no symmetries and
is therefore probably the smallest non-integrable physical system.
\end{abstract}


\maketitle

\section{}

The Rabi 
(or single-mode spin-boson) model 
constitutes
 probably the simplest physical system beyond the
harmonic oscillator.
Introduced over 70 years ago \cite{rabi}, 
its
applications range from quantum optics 
\cite{vedral} 
and 
magnetic resonance 
to solid
state \cite{irish} and molecular physics 
\cite{thano}. 
Very recently, it has gained a prominent role
in novel fields of research such as cavity QED 
\cite{engl}
and circuit QED \cite{niemc}. 
It can be experimentally realized in Josephson junctions
\cite{sorn} or using trapped ions
\cite{leib}, in 
Cooper-pair boxes \cite{wallr}
and flux q-bits \cite{forn}. In this way, its complete
theoretical understanding is mandatory for all feasible
approaches to quantum computing \cite{pell}.
Despite its old age and central importance, the Rabi model has not
been
exactly solved \cite{irish,larson,werl}. 
It shares with the other
paradigma of quantum physics, the hydrogen atom, an
infinite-dimensional 
state
space
but --- in contrast to the latter ---
the spectrum and eigenfunctions of the Rabi model are known only 
by numerical diagonalization in a truncated, finite-dimensional  Hilbert space.
This is quite surprising, as the Rabi model has a smaller number of
degrees of freedom 
than
the hydrogen atom. In particular, a single degree of
freedom $\hat{x}$, subject to a harmonic potential, couples to a
quantum system with only two allowed states $|\ua\rangle$ and
$|\da\rangle$. Therefore, it does not possess a classical limit:
the quantum degree of freedom has a finite-dimensional Hilbert space and places
the Rabi model in between the  case of one 
and two (classical) degrees of freedom. 
The hamiltonian reads
 in units with $\hbar=1$,
\beq
H_R= \om\ad a +g\s_x(a+\ad) +\D\s_z. 
\label{ham1}
\eeq
Here, the $\s_{x,z}$ are Pauli matrices, describing the two-level
system with level-splitting $2\D$ and $a$ ($\ad$) denote
destruction (creation) operators of a single bosonic mode with
frequency $\om$. These two systems are coupled through a term
proportional to $g$, which has different interpretations according to
the experimental situation to model. 
Although (\ref{ham1}) 
describes
the simplest of all 
physically sensible interacting
quantum systems, it poses a serious obstruction to its analytical solution
because of the apparent lack of a second conserved quantity besides
the energy, which
has 
led to the widespread opinion that it cannot be
integrable 
\cite{mil,gra,kus,bonci,fukuo,emar}.
 To remedy this difficulty,
Jaynes and Cummings 
proposed already in the sixties 
an approximation to (\ref{ham1}) which does possess such a
quantity 
\cite{jaynes}.
Their hamiltonian reads 
\beq
H_{JC} = \om\ad a +g(\s^+a + \s^-\ad) +\D\s_z.
\label{JC}
\eeq
with $\s^\pm = (\s_x\pm i\s_y)/2$.\\
Here, the operator $C=\ad a +\frac{1}{2}(\s_z+1)$ commutes with
$H_{JC}$ and leads
at once to the solvability of (\ref{JC}). The
Jaynes-Cummings model is the
so-called ``rotating-wave'' approximation 
to (\ref{ham1}) and
was justified because the
conditions of
near-resonance $2\D\approx\om$ and  weak
coupling $g\ll \om$ for such an approximation
are realized in many experiments.
The conservation of $C$ means that the state
space
decomposes into an infinite sum of two-dimensional invariant
subspaces,
each labeled by the value of $C=0,1,2,\ldots$ Each eigenstate of
(\ref{JC}) is then characterized by two quantum numbers, namely
$C$ and a two-valued index, for example $+$ and $-$, denoting a basis vector
in the two-dimensional subspace which belongs to $C$. Whereas the
possible values of $C$ form an unbounded set, corresponding to the
quantization of a classical degree of freedom, the second quantum
number
can take only two values, reflecting the intrinsic quantum
nature of the two-level system. 
The conserved quantity $C$ generates a continuous $U(1)$-symmetry  
of the Jaynes-Cummings model (\ref{JC}) which is broken down to $\Zz_2$
in the Rabi model (\ref{ham1}) due to the presence of the term $\ad\s^++a\s^-$.
This residual $\Zz_2$-symmetry, usually called parity, leads to a
decomposition of the state space into just two subspaces ${\cal{H}}_\pm$, each
 with infinite dimension. One would conclude that this symmetry cannot
suffice to solve the model exactly --- but in fact it does. 
Whereas a discrete symmetry is too weak to accommodate a classical
(continuous) degree of freedom, it can do so with a quantum 
degree of freedom --- like the two-level system in the Rabi model.

We observe a direct relation between the nature of the degree of freedom
(continuous or discrete) and the symmetry ($U(1)$ versus $\Zz_2$),
which can be used to ``eliminate'' it by fixing the corresponding 
irreducible representation.  

\begin{figure}[ht]
\begin{center}
\includegraphics[width=0.9\columnwidth,clip]{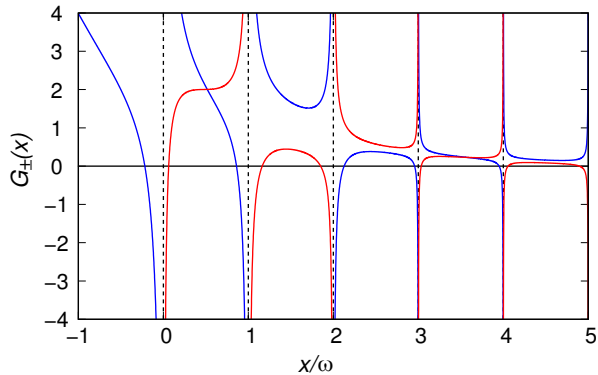}
\end{center}
\caption{$G_+(x)$ (red) and $G_-(x)$ (blue) 
in the intervall $[-1,5]$ for $g=0.7$, $\om=1$ and
  $\D=0.4$} 
\label{G-res-full}
\end{figure}

Our main result is the following: 
The spectrum of (\ref{ham1}) consists of two parts, the
{\it regular} and the {\it exceptional} spectrum. Almost all
eigenvalues are regular and given by the zeroes of the 
transcendental function 
$G_\pm(x)$ in
the variable $x$, which is defined 
through its power series in the coupling $g$:
\beq
G_\pm(x)=\sum_{n=0}^\infty K_n(x)
\left[1\mp\frac{\D}{x-n\om}\right]\left(\frac{g}{\om}\right)^n.
\label{sol}
\eeq  
The coefficients $K_n(x)$ are defined recursively,
\beq
nK_{n}=f_{n-1}(x)K_{n-1}-K_{n-2},
\label{recur}
\eeq
with the initial condition $K_0=1, K_1(x)=f_0(x)$
and
\beq
f_n(x)=\frac{2g}{\om}+\frac{1}{2g}
\left(n\om-x +\frac{\D^2}{x-n\om}\right).
\label{f-n}
\eeq
The function $G_\pm(x)$ is not analytic in $x$
but has simple poles for $x=0,\om,2\om,\dots$,
 these poles are precisely the eigenvalues
of the uncoupled bosonic mode. Then the regular energy spectrum of the
Rabi model in each invariant subspace ${\cal H}_\pm$ with parity $\pm 1$
is given by the zeroes of $G_\pm(x)$: for all values $x^\pm_n$ with
$G_\pm(x^\pm_n)=0$, the $n$-th eigenenergy with parity $\pm 1$ 
reads $E_n^\pm=x^\pm_n-g^2/\om$. The functions $G_\pm(x)$ are plotted
in Fig.\ref{G-res-full} for  $\D=0.4$ 
and $\om=g=1$ between
$x=-1$ and $x=5$. Their zeroes determine the first six
eigenenergies with parity $+1$ and the first five (including the
groundstate) with parity $-1$.
 For special values of model parameters $g,\D$, there are
eigenvalues which do not correspond to zeroes of (\ref{sol}), these
are the exceptional ones. All exceptional eigenvalues have the form 
$E^e_n=n\om-g^2/\om$, that is, they lie on one of the so-called baselines 
\cite{quasi-exact} and coincide with some point in the spectrum of the 
limiting case $\D=0$, which corresponds in spin-boson language to
zero hybridization. All exceptional eigenvalues are doubly
degenerate
with respect to parity. The necessary and sufficient condition for an
exceptional eigenvalue to lie on the $n$-th baseline reads,
\beq
K_n(n)=0,
\label{except}
\eeq
which furnishes a condition on the model parameters $g$ and $|\D|$.
As all the $K_n(n)$ are independent polynomials in $\D^2$, there can
exist
at most two exceptional eigenvalues for given $g,\D$. These
exceptional solutions to (\ref{ham1}) have been known for a long time
and
were first discovered by Judd \cite{quasi-exact}. 
They occur when the pole of $G_\pm(x)$ at $x_n=n\om$ is lifted
because its numerator in (\ref{sol}) vanishes which happens only if
(\ref{except}) is satisfied.

\begin{figure}[h]
\begin{center}
\includegraphics[width=0.9\columnwidth,clip]
{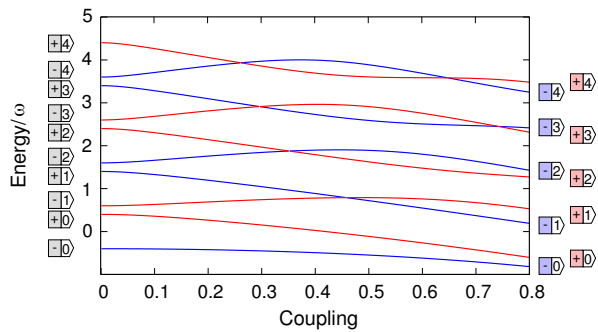}
\end{center}
\caption{\small Rabi spectrum for $\D=0.4$, $\om=1$ and $0\le g \le 0.8$ in the
  spaces
with positive (red) and negative (blue) parity. Within each space
the states are labeled with ascending numbers $0,1,2,\ldots$. This
labeling does not change with $g$ because no lines intersect within
spaces of fixed parity. However, level crossings
of states with different parity occur. The spectral graph is composed
of two intersecting ``ladders'' of level lines, each corresponding to
one parity subspace. This labeling is used on the right side of the
figure. On the
left side the states with $g=0$ are labeled by the uncoupled degrees
of freedom, i.e. in $|\pm,n\rangle$, 
$+/-$ corresponds to the two-level system and
$n=0,1,2,\ldots$
to the eigenstates of the bosonic mode. 
 } 
\label{spec1}
\end{figure}

The functional form of
$G_\pm(x)$ reads,
\beq
G_\pm(x) = G_\pm^0(x) +\sum_{n=0}^\infty\frac{h^\pm_n}{x-n\om},
\label{GG0}
\eeq 
where $G^0_\pm(x)$ is entire in $x$. 
The position of the solutions to $G_\pm(x)=0$
is dictated by the pole-structure of $G_\pm(x)$, 
which leads to the
conjecture that the number of eigenvalues in each interval $[n\om,(n+1)\om]$
is restricted to be 0, 1, or 2. Moreover, for large energies 
$x\gg g,\D$, the
entire part of $G_\pm(x)$ can be approximated by
\beq
G_\pm^0(x)=\left(1\mp\frac{\D}{x}\right)\exp\left(-\frac{x}{2\om}\right) 
\eeq
which is monotonous, always $>0$ and slowly varying on the 
scale set by $\om$. 
This suggests that the relative position of the solution
$x_j(n)$ in the vicinity of the $n$-th baseline is fixed by the sign
of $h^\pm_n$ alone: If $h^\pm_n>0$ ($h^\pm_n<0$) then $x_j(n)$ lies
in the
interval $[(n-1/2)\om,n\om]$ ($[n\om,(n+1/2)\om]$). It would also entail that an
interval $[n\om,(n+1)\om]$ with two roots of $G_\pm(x)=0$ can 
only be adjacent to an interval
with one or zero roots; in the same way, an empty interval can never
be adjacent to another empty interval. 
These conjectures about the
distribution
of the large eigenvalues, 
which can be
confirmed numerically,
lead to a fairly regular distribution of the energies
and a violation of the Berry-Tabor criterion \cite{ber-tab,kus}.
Fig.\ref{spec1} shows the lowest part of the spectrum
for $\D=0.4$, $\om=1$ and $g$ between $0$ and $1$. 
One may prove that
there are no level crossings within each parity subspace, allowing the
unique labeling of each state $|\psi\rangle$ with a pair of two quantum
numbers, $|\psi\rangle=|n_0,n_1\rangle$:
The parity quantum number
 $n_0$, which takes the values $+1$ and $-1$, and
$n_1=0,1,2,\ldots$, which denotes the $n_1$-th zero of $G_{n_0}(x)$.
The exceptional solutions correspond to level crossings
between $\Hp$ and $\Hm$.
This characterization of each eigenstate through two quantum numbers
corresponding to the degrees of freedom of the system parallels
the unique assignment of three quantum numbers $n,l,m$ to the
eigenstates of the hydrogen atom, reflecting the quantization of
radial and angular degrees of freedom, a hallmark of integrability.

It seems therefore natural to call a quantum system integrable when such
an assignment can be made --- {\it independent} of the explicit
determination of conserved quantities or even action variables, which
is only possible if the system under consideration has an integrable
classical limit in the sense of Liouville \cite{arnold}.
Without making the assumption of a classical limit, our criterion
reads,\par
\vspace{3mm}
\noindent
{\it Criterion of quantum integrability:}\ \ If each
eigenstate of a quantum system with
$f_1$ discrete and $f_2$ continuous degrees of freedom can be {\it uniquely}
labeled by $f_1+f_2=f$ quantum numbers 
$\{d_1,\ldots,d_{f_1},c_1,\ldots c_{f_2}\}$, 
such that the $d_j$ can take on dim$({\cal H}_j)$ different values,
where ${\cal H}_j$ is the state space of the $j$-th discrete 
degree of freedom   
and the $c_k$ range from zero to infinity, then this system is 
quantum integrable.\par
\begin{figure}[ht]
\begin{center}
\includegraphics[width=0.9\columnwidth,clip]
{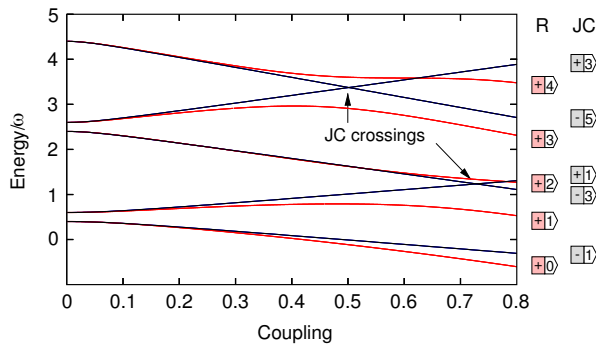}
\end{center}
\caption{\small The spectrum of the Jaynes-Cummings model (black) compared with the
Rabi model (red) for $\D=0.4$ and even parity. 
The state labeling of the former has the form
$|\psi\rangle=|n_0,C\rangle$ with $n_0=\pm $. Two accidential crossings
occur between levels with $C=5$ and $C=3$ at $g\approx 0.5$ and
between $C=1$ and $C=3$  at $g\approx 0.73$.
} 
\label{rwa}
\end{figure}

\vspace{3mm}
\noindent
The criterion does not presuppose the existence of a family of
commuting
operators whose different spectra are associated with the $\{d_j,c_k\}$
but provides a condition on the spectral graph of the system, that is,
the spectrum as function of a parameter, typically one of the coupling
constants.
Without such a deformation parameter (which must conserve integrability)
the association of more than one quantum number to the levels of a
non-degenerate spectrum is ill-defined and would be restricted
either to models solvable via Bethe ansatz \cite{natan} or
systems with integrable classical limit.
As is well-known, the Berry-Tabor criterion \cite{ber-tab} relies
precisely on the existence of this limit, that is, it applies
only to the correspondence limit of large quantum numbers and
presumes the validity of the semi-classical quantization scheme.
The criterion proposed here is not restricted to the upper part
of the spectrum or the existence of a classical limit.
Moreover, it can be given 
a phenomenological formulation using the spectral graph: 
If the system with
coupling $g$ has
$f>1$ discrete or continuous 
degrees of freedom, each eigenstate can be written as
 $|n_1,\ldots n_f;g\rangle$.
Upon varying the
coupling $g$, each of these states defines
an energy  level $E(n_1,\ldots n_f;g)$ as function of $g$. 
The totality of states for all sets $\{n_1,\ldots n_f\}$ 
forms an $f$-dimensional manifold
of spectral lines, which will typically intersect each other when
energies corresponding to different invariant subspaces become
accidentally degenerate at special values of the coupling. 
In a fully nondegenerate spectrum, a single integer quantum number which
assigns $0,1,2,\ldots$ to the eigenstates with ascending energies would
suffice for a unique description of each state --- but whenever a
level
crossing occurs, this is no longer possible. A second quantum number
is
necessary to discern the energetically degenerate states. The crossing
is
{\it accidental} in the sense that no new symmetry 
appearing at the given value of the coupling parameter is responsible for
the degeneracy, which will concern typically only two states in the
spectrum.
It is merely the fact that for these states, belonging to dynamically
decoupled subspaces, the energy dependence on the coupling coincides.
The second quantum number labels the invariant subspaces and can be
used to discern the energetically degenerate states at the crossing
point. According to the proposed criterion, Integrability is equivalent
to the existence of $f$ numbers to classify energy levels uniquely, 
if the system has $f$ continuous
or discrete degrees of freedom. It should be emphasized that these
``quantum numbers'' are a more general concept than the radial and
angular quantum numbers known from atomic physics --- they need not 
correspond to physical quantities (actions) quantized in integer multiples of
$\hbar$ and have nothing to do with the Bohr-Sommerfeld semi-classical
quantization rules. They are only defined with respect to the
unique description of eigenstates; the integrable systems
differ from the non-integrable ones because they allow for a ``fine-grained''
description through an $f$-dimensional vectorial label, 
whereas the latter have a one-dimensional label corresponding to
energy as the only conserved quantity.      
In the Rabi model we have $f_1=f_2=1$ and  
degeneracies takes place
between levels of states having
different parity, whereas within the parity subspaces no level
crossings occur. 
The spectral graph consists of two ``ladders'' of level lines
$|n_0,m\rangle$ for $m=0,1,2,\ldots$. Each
ladder corresponds
to an invariant subspace of the $\Zz_2$-symmetry
characterized by $n_0=\pm$, the parity eigenvalue. Within each subspace
the system corresponds to a single continuous degree of freedom and
is therefore integrable by definition. The global label (valid for all
values of the coupling $g$) is two-dimensional as $f=f_1+f_2=2$;
the
Rabi model belongs therefore to the class of integrable systems. 
It may be, however, that the symmetry is even
stronger
than necessary to achieve integrability, analogous to the classically
``superintegrable'' systems \cite{evans}.      
The Jaynes-Cummings model
is an example for this case.
Here, the continuous $U(1)$-symmetry leads to a further
decomposition of the subspaces with fixed parity into a direct
sum of two-dimensional invariant spaces labeled by the
unbounded quantum number $C=0,1,2,\ldots$ ; even (odd) values of
$C$ correspond to odd (even) parity. The larger number of dynamically
decoupled state spaces entails a second possibility to label the states 
uniquely: through $C$ and a two-valued index $n_0=\pm$. Now, all levels with
different $C$ may intersect, leading not to just two but infinitely
many
ladders labeled by $C$, each having two rungs, labeled by $n_0$.
In Fig.\ref{rwa} the four lowest levels of the Rabi model with positive parity
(red lines) 
are compared with the corresponding levels of the Jaynes-Cummings
model
with $C=1,3,5$ (black lines). 
The enlarged symmetry of the latter leads to  two level
crossings which are not present in
the Rabi model \cite{stepan}.
The appearence of intersecting ladders in the spectral graph can be
detected without knowledge of the exact solution or the correct
assignment of quantum numbers to the different levels, the only
condition
being a sufficient numerical resolution to discern degeneracies from
narrow avoided crossings. This is a phenomenological virtue of the
proposed criterion which could be used in computer experiments to
test whether a given numerically solvable system possesses a hidden
integrable structure. 
Although a large number of level crossings as a function of  model
parameters gives a strong hint to integrability, it is
difficult to make the argument 
quantitative,
because the number of 
intersecting ladders could become infinite already for two continuous
degrees of freedom. 
For systems within the present class of one continuous and one discrete
degree of freedom, however, its application is fairly obvious.
The proposed criterion is sufficient but not necessary: in systems
with factorized scattering matrices, a unique association of eigenstates 
with quantum
numbers is possible which treats the discrete degrees of freedom
differently from the present scheme, via the so-called nested Bethe ansatz
 \cite{natan}. But also 
this labeling leads to intersecting ladders   
as function of the coupling parameter.
 
\begin{figure}[ht]
\begin{center}
\includegraphics[width=0.9\columnwidth,clip]{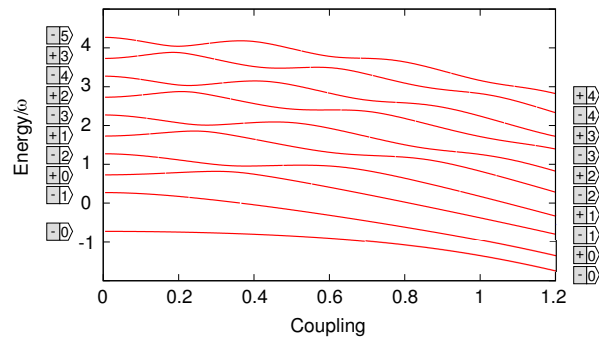}
\end{center}
\caption{\small The spectrum of the generalized Rabi model with broken
$\Zz_2$-symmetry and $\Delta=0.7, \eps=0.2$. The two-fold labeling of the states
corresponds to the two integrable limits $g=0$ on the left
and $g\rightarrow\infty$ on the right. For finite $g$ neither labeling
classifies the states properly.} 
\label{z2}
\end{figure}

On the other hand, the {\it absence} of any level crossings in the
spectral graph is sufficient for non-integrability if the total number
of degrees of freedom (continuous and discrete) 
exceeds one: it means that the states can be
classified only by energy, the single conserved quantity always
present in hamiltonian systems, and no invariant subspaces
exists.
In the context of the Rabi model this case can be realized by
breaking the $\Zz_2$-symmetry. A possible generalization of
(\ref{ham1}) reads
\beq
H_\eps= \om\ad a +g\s_x(a+\ad)+\eps\s_x +\D\s_z. 
\label{hamz}
\eeq
The term $\eps\s_x$ breaks the parity symmetry which couples
the bosonic mode and the two-level system. 
Physically it corresponds to a spontaneous transition of the two-level
system which is not driven by the radiation field.
The state space does not
separate
into two subspaces and indeed the spectral graph exhibits no level
crossings at all if $\eps$ is not a multiple of $\om/2$ \cite{om2}. 
This is shown in Fig.\ref{z2}.
In this situation, the eigenstates can be uniquely numbered as belonging to the
$n$-th energy level in ascending order, $|\psi\rangle=|n\rangle$.
We have only one quantum number, energy, corresponding to the sole
conserved quantity. Because the number of degrees of freedom 
nevertheless exceeds
one, this model must be considered non-integrable.
Interestingly, (\ref{hamz}) is still exactly solvable, although it does
not possess any symmetry. Define the functions
\beq
R^\pm(x)=\sum_{n=0}^\infty K^{\pm}_n(x)\left(\frac{g}{\om}\right)^n
\eeq 
and
\beq
{\bar{R}}^\pm(x)=
\sum_{n=0}^\infty \frac{K^{\pm}_n(x)}{x-n\om\pm\eps}\left(\frac{g}{\om}\right)^n.
\label{z2br}
\eeq
The $K^\pm_n(x)$ are again recursively defined,
\beq
nK^\pm_{n}=f^\pm_{n-1}(x)K^\pm_{n-1}-K^\pm_{n-2},
\label{recur-z2}
\eeq
with the initial condition $K^\pm_0=1, K^\pm_1(x)=f^\pm_0(x)$
and
\beq
f^\pm_n(x)=\frac{2g}{\om}+\frac{1}{2g}
\left(n\om-x\pm\eps +\frac{\D^2}{x-n\om\pm\eps}\right).
\label{f-n-z2}
\eeq
The $n$-th eigenvalue $E_n$ of (\ref{hamz}) is given by the 
$n$-th zero $x_n$
of
\beq
G_\eps(x)=\D^2{\bar{R}}^+(x){\bar{R}}^-(x)-R^+(x)R^-(x) 
\eeq
through $E_n=x_n-g^2/\om$.
The fact that $H_\eps$ can be diagonalized analytically although 
not even a discrete symmetry is present signifies that
integrability and solvability are not equivalent in the realm of
quantum physics. In contrast to classical mechanics,  
non-integrable quantum 
systems with exact solutions exist. 
\section{Conclusions}
We have seen that the Rabi model is integrable contrary to common
belief. The exact spectrum is given by the zeroes of a well-defined
transcendental function $G_\pm(x)$ which converges everywhere except
at the set of points $x\in\{n\om|n=0,1,2,\ldots\}$, where it has
simple poles. The previously found exact eigenvalues for special
values of the parameters $g,\D$ are always two-fold degenerate
 and correspond to the merging of a zero 
with 
a pole of $G_\pm(x)$ at $x=n\om$.
The eigenfunctions can be given explicitely as well 
(see appendix).
The solution of the Rabi model appears in this way to be of comparable
complexity with the well-known solution of the one-dimensional
potential well, although it is a strongly interacting, fully quantum
mechanical model. Its generalization with broken $\Zz_2$-symmetry
is non-integrable according to our criterion on integrability,
which
is tailored to the present class of systems with more than one but
less than two degrees of freedom - nevertheless it is exactly solvable
and describes 
probably
the smallest non-integrable physical system.
Whether exact solutions are possible for 
non-integrable quantum models with two or more continuous degrees of freedom
is under current investigation. 
\section{Appendix}
To derive the result 
(\ref{sol})-(\ref{f-n}) for $G_\pm(x)$
we use the
representation of bosonic creation and anihilation operators
in the Bargmann space of analytical functions in a complex variable
$z$ \cite{bargmann},
\beq
a \rightarrow \frac{\partial}{\partial z}, \quad \ad \rightarrow z
\eeq
Then, after applying a Fulton-Gouterman transformation
and setting $\om=1$, the
time-independent
Schr\"odinger equation in the subspace $\Hp$ with even parity reads
\beq
z\dd\psi(z) +g\left(\dd + z\right)\psi(z) = E\psi(z) -\D\psi(-z),
\label{time-i-schr}
\eeq
which is a {\it functional} differential equation in 
$z$. The solution $\psi(z)$ is analytic in the whole complex
plane
if $E$ belongs to the spectrum of $H_R$.
 With the notation
$\psi(z)=\p_1(z)$, $\psi(-z)=\p_2(z)$, one obtains a coupled system
of first-order equations,
\beq
\begin{array}{lll}
(z+g)\dd\p_1(z) +(gz-E)\p_1(z) +\D\p_2(z) &=& 0\\
(z-g)\dd\p_2(z) -(gz+E)\p_2(z) +\D\p_1(z) &=& 0. 
\end{array}
\label{coup-sys}
\eeq
With $y=z+g$, $x=E+g^2$, $\p_{1,2}=e^{-gy+g^2}\bp_{1,2}$,
it follows,
\begin{eqnarray}
\phantom{(y-2)}y\frac{\rd}{\rd y}\bp_1 &=& x\bp_1-\D\bp_2
\label{coup-sys2}\\
(y-2g)\frac{\rd}{\rd y}\bp_2 &=& (x-4g^2+2gy)\bp_2 -\D\bp_1.
\label{coup-sys3}
\end{eqnarray}
$\bp_2(y)$ is expanded into a power series in $y$,
\beq
\bp_2(y)=\sum_{n=-\infty}^\infty K_n(x)y^n 
\eeq
and from (\ref{coup-sys2})
one obtains for $\bp_1(y)$,
\beq
\bp_1(y) = \sum_{n=-\infty}^{\infty}K_n(x)\frac{\D}{x-n}y^n.
\label{phi1}
\eeq
Eq.(\ref{coup-sys3}) is then equivalent with the recurrence
(\ref{recur}) and (\ref{f-n}). For $\bp_{1,2}(y)$ to be analytic in 
$y$ (i.e. at the point $z=-g$) 
it is necessary that $K_n(x)=0$ for all $n<0$. If the
coefficient $K_0$ is set to 1, this entails $K_1(x)=f_0(x)$, which
fixes
the initial conditions of (\ref{recur}). However, the power series in 
$y$, though analytic at $z=-g$, has the finite radius of convergence
$R=2g$, which can be deduced from the
asymptotic value $1/(2g)$ of $f_{n-1}(x)/n$ for $n\rightarrow\infty$.
Therefore, $\bp_2(z+g)$ will develop a branch-cut at $z=g$, if
the
parameter $x$ does not belong to the discrete subset of $\R$ which
determines the spectrum of $H_R$ in $\Hp$. 
 But this condition on $x$ follows
easily from the fact that we have {\it two} representations for
$\psi(z)$ in $\Hp$, one constructed via $\p_2(z)$ and the other with
$\p_1(z)$,
\begin{eqnarray}
\psi(z)=\p_2(-z)&=& e^{gz}\sum_{n=0}^\infty K_n(x)(-z+g)^n \label{psi-2}\\
\psi(z)=\p_1(z) &=& e^{-gz}\sum_{n=0}^\infty K_n(x)\D\frac{(z+g)^n}{x-n}.
\label{psi-1}
\end{eqnarray}
The expansion in powers of $z$ is analytic at $z=g$ in (\ref{psi-2})
and analytic at $z=-g$ in (\ref{psi-1}). Only if both expansions are
analytic at $z=\pm g$ they may coincide everywhere and represent the same
$\psi(z)$ for all $z\in \Cc$. Therefore
\beq
G_+(x;z)= \p_2(-z)-\p_1(z) =0\ \ {\textrm{for all}}\ \ z\in\Cc  
\label{G-eq}
\eeq  
if and only if $x=E+g^2$ corresponds to a point in the spectrum of $H_R$.
Because $x$ is the only variable in $G_+(x;z)$ besides $z$, it
suffices to solve $G_+(x;z)=0$ for some arbitrarily chosen
$z$. However, the function $G_+(x;z)$ is only well-defined in $x$ via its
expansion in powers of $z$ within the joint radius of convergence of
(\ref{psi-2}) and (\ref{psi-1}), which restricts the absolute value
of $z$ to be less then $g$. If $x_0$ solves $G_+(x;z)=0$
for one such $z$, it will solve (\ref{G-eq}) as well and the
eigenfunction  $\psi(z;E)$ for the energy $E=x_0-g^2$ possesses the two
alternative
series
expansions (\ref{psi-2},\ref{psi-1}) which converge
for $x=x_0$ in the whole $z$-plane and are uniquely determined by
$x_0$. It follows that all regular energy levels of $H_R$ 
with positive parity are
non-degenerate. Setting now $z=0$, we obtain $G_+(x)=G_+(x;0)$.
This argument for $\Hp$
carries over to $\Hm$, the subspace 
with negative parity, by replacing $\D$ with $-\D$, which completes the
derivation of (\ref{sol}).
An exceptional solution  occurs when a zero of
$G_\pm(x)$ merges with a pole at $x=n\om$. 
As then $\p_2(-z)\neq\p_1(z)$, this
point corresponds to a two-dimensional representation of $\Zz_2$
and the eigenvalue is two-fold degenerate.

There have been early attempts \cite{schweb,swain,durst}
to obtain an equation for the eigenvalue $E(x)$ 
without the
$\Zz_2$-symmetry underlying the functional equation
(\ref{time-i-schr}), as (\ref{coup-sys}) arises
after a rotation in spin space directly from (\ref{ham1}).
Then, of course, no functional relation between 
$\p_1$ and $\p_2$ can be used.
Schweber has found the recurrence (\ref{recur},\ref{f-n})
for the coefficients $K_n(x)$ in
\beq
\bp_2(y)=\sum_{n=0}^\infty K_n(x)y^n
\label{expan}
\eeq
and noted that the solution belongs only to the Hilbert space if
a branchcut at $y=2g$ is avoided by the correct choice of $x$.
One may formulate the problem in terms of the two kinds of
solutions to the three-term recurrence (\ref{recur})
with initial condition $K_0=1$. The {\it dominant}
solutions \cite{gautschi} all have the asymptotics
\beq
\lim_{n\ra\infty}\frac{K_n(x)}{K_{n-1}(x)}=\frac{1}{2g}=\frac{1}{R},
\label{lim}
\eeq
with finite radius of convergence $R=2g$ of (\ref{expan})
whereas the {\it minimal} solution has $\lim_{n\ra\infty}K_n(x)/K_{n-1}(x)=0$,
which entails an infinite radius of convergence of (\ref{expan}), 
as mandated by analyticity. 
This imposes a second constraint on the initial condition for
(\ref{recur}) besides 
\beq
K_0=1,\quad K_1(x)=f_0(x),
\label{initial}
\eeq
i.e. on the
allowed value of $K_1$, 
because the minimal solution is unique, whereas the
dominant solutions are only determined modulo a multiple of the
minimal one. 
This second constraint fixes $x$ and therefore the spectrum.
The procedure is then the following:\\
1.) Choose an arbitrary value $x\in\mathbb{R}$.\\
2.) Compute numerically the minimal 
solution to (\ref{recur}) with $K_0=1$,
$\{K_n^{min}(x), n=1,2,\ldots\}$. The minimal solution depends
 on the $f_n(x)$ with $n\ge 1$.\\
3.) Compare $K_1^{min}(x)$ with $f_0(x)$.\\
If both coincide, $x$ will correspond to the eigenenergy $E=x-g^2$.

In \cite{schweb,swain} it is claimed that a formally exact 
solution of the
eigenvalue problem may be thus obtained in the following way: 
Define $V_n=K_n/K_{n-1}$ for $n=1,2,\ldots$ Then (\ref{recur}) 
with initial condition (\ref{initial}) is
equivalent to the two-term nonlinear recurrence relation
\beq
V_n=\frac{f_{n-1}(x)}{n} - \frac{1}{nV_{n-1}}, \quad n \ge 2
\label{recur2}
\eeq
with initial condition $V_1=f_0(x)$, if none of the $K_n$ vanishes.
But (\ref{recur2}) can also be written as
\beq
V_{n}=\frac{1}{f_{n}(x) -(n+1)V_{n+1}}
\label{recur3}
\eeq
which determines $V_n$ ``from above'' instead by the initial condition
$V_1=f_0(x)$ together with (\ref{recur2}). One could write $V_1$ as a
continued fraction,
\beq
V_1(x)=\frac{1}{f_1(x)-\frac{2}{f_2(x)-\frac{3}{f_3(x)-\ldots}}}.
\label{cont-frac} 
\eeq
Because the r.h.s. of (\ref{cont-frac}) depends only on the set
$f_1(x),f_2(x),\ldots$, the formal equation
\beq
f_0(x)=1|f_1(x)-2|f_2(x)-3|f_3(x)-\ldots
\label{cont-frac2}
\eeq
seems to provide the sought condition on $x$, a transcendental
equation, whose infinitely many roots determine the spectrum of $H_R$. 
This argument, however, is fallacious, because (\ref{cont-frac2}) is
only a reformulation of the recurrence relation (\ref{recur2}) with
initial condition $V_1(x)=f_0(x)$ and therefore valid for {\it all}
$x$, regardless, whether they lie in the spectrum or not. Indeed,
one may rewrite (\ref{cont-frac2}) as the finite continued fraction
\beq
f_0(x)=1|f_1(x)-2|f_2(x)-\ldots (n-1)|f_{n-1}(x)-nV_n(x),
\eeq
where $V_n(x)$ depends on $f_0(x)$ via (\ref{recur2}) and amounts then
simply to
\beq
f_0(x)=f_0(x).
\label{id}
\eeq
Only if the r.h.s of (\ref{cont-frac2}) is {\it approximated} in one
way or the other, it may give a non-trivial equation for $x$, whereas
its
{\it exact} value is tantamount to (\ref{id}). In other words, the
continued fraction in (\ref{cont-frac},\ref{cont-frac2}) 
has no well-defined value without specifying
the limiting behavior of the $V_n$ for $n\ra\infty$. It will converge
both for the dominant solutions with $V_n\ra 1/2g$ as well as for
the sought minimal solution with $V_n\ra 0$, the difference being that
in the latter case the resulting $V_1$ will not depend strongly on the
cut-off procedure, i.e. setting $V_N=\veps$ with $\veps\approx 0$ for
some sufficiently large $N$, whereas the choice of $N$ and $\veps$
has a strong impact on $V_1$, if one cuts off the continued
fraction with $V_N=\veps+1/2g$ \cite{gautschi}.
This, of course,
follows
from
the uniqueness of the minimal solution to (\ref{recur}). 

Therefore (\ref{cont-frac2}) is equivalent to the numerical algorithm
described above. Moreover, the cut-off at $n=N$, necessary  
to define its r.h.s., renders it equivalent with a diagonalization of
$H_R$ in a finite-dimensional Hilbert space, whose dimension depends on $N$.

Clearly, this problem with the representation of the spectral condition on $x$
in terms of an infinite continued fraction is a consequence of 
neglecting the $\Zz_2$-symmetry which relates $\p_1(z)$ and $\p_2(z)$ 
in a nontrivial way and allows the definition of the function
$G_\pm(x)$, whose zeroes determine the energy eigenvalues of the
Rabi hamiltonian.
\begin{acknowledgments}

 I wish to thank K.H. H\"ock, N. Andrei, 
M. Dzierzawa, S. Graser and T. Kopp for stimulating discussions.
This work was supported by the DFG through TRR80.
\end{acknowledgments}


\begin{thebibliography}{99}
\bibitem{om2} If $\eps$ is an integer multiple of $\om/2$, the model
becomes integrable again and concomitant level crossings appear.
\end{thebibliography}
\end{document}